\documentstyle[preprint,revtex]{aps}
\begin{document}
\hfill\vbox{\hbox{\bf NUHEP-TH-93-27}\hbox{October 1993}}\par
\thispagestyle{empty}
\begin{title}
\begin{center}
{\bf Possibility of Studying Electroweak Symmetry Breaking at {\boldmath
$\gamma \gamma$} Colliders}
\end{center}
\end{title}
\author{Kingman~Cheung}
\begin{instit}
Dept. of Physics \& Astronomy, Northwestern University, Evanston,
Illinois 60208, USA\\
\end{instit}
\begin{abstract}
\nonum
\section{Abstract}
The studies of the electroweak symmetry breaking sector (EWSBS) at
 $\gamma\gamma$ colliders were considered previously in the loop processes of
$\gamma\gamma \to w_Lw_L,\,z_Lz_L$, but they are suffered from the huge
$W_T W_T$ and $Z_TZ_T$ backgrounds.  Here we present another possible process
  that involves spectator $W$'s and $W_L$'s, the latter of
which are scattered strongly by the interactions of the EWSBS.
We also show that this process should be safe from the
transverse backgrounds and it can probe the structure of the EWSBS.
\end{abstract}
\newpage

\section{Introduction}
\label{intro}

So far very little is known about the electroweak
symmetry-breaking-sector (EWSBS),
except it gives  masses to the vector bosons via the spontaneous symmetry
breaking, and masses to fermions via the Yukawa couplings.
In the minimal standard model (SM) one scalar Higgs boson is responsible for
the electroweak symmetry breaking but its mass is not determined by the model.
If in the future no Higgs boson is found below 800 GeV,
the heavy Higgs scenario ($\approx 1$ TeV) will imply a strongly interacting
Higgs sector because  the Higgs self-coupling $\lambda\sim m_H^2$
becomes strong \cite{quigg}.  However, there is no evidence to favor the
model with a scalar Higgs, and
so any models that can break the electroweak symmetry the same way as the
single Higgs does can be a candidate for the EWSBS.

One of the best ways to uncover the underlying dynamics of the EWSBS is to
study the longitudinal vector boson scattering \cite{quigg,chano}.
The Equivalence Theorem (ET) recalls, at high energy,
 the equivalence between the longitudinal part $(W_L)$ of the vector bosons
to the corresponding Goldstone bosons ($w_L$) that were ``eaten"
in the Higgs mechanism.  These Goldstone bosons
originate from the EWSBS so that their scattering must be via the interactions
of the EWSBS, and therefore the $W_L W_L$ scattering can reveal the dynamics
of the EWSBS.

The strong $W_L W_L$ scattering
have been studied quite seriously at the hadronic
supercolliders \cite{bagger}, but less at the $e^+e^-$ colliders,
 and very little at the $\gamma\gamma$ colliders.
In hadronic colliders, only the ``gold-plated" modes, the leptonic decays of
the $W$ and $Z$ bosons, have been considered due to the messy hadronic
backgrounds; whereas in $e^+e^-$ and $\gamma\gamma$ colliders
one can make use of the hadronic decay mode or mixed decay mode of the final
state $W$'s or $Z$'s.
With the advance in the photon collider designs it
is possible to construct an almost monochromatic $\gamma\gamma$ collider
based on the next generation linear $e^+e^-$ colliders using the laser
backscattering method \cite{teln}.
The monochromaticity of the photon beams depends on the polarizations of the
initial electron and the laser photon.  The polarizations of the
initial electron and the laser photon can be  adjusted to maximize the
monochromaticity of the photon beam \cite{teln} with a center-of-mass
energy about 0.8 of the parent $e^+e^-$ collider.  Hence, a
 2 TeV $e^+e^-$ collider will
give a 1.6 TeV $\gamma\gamma$ collider by the laser backscattering method.
For the following we will assume a monochromatic $\gamma\gamma$ collider of
energy 1.5 TeV with an integrated luminosity of 100~fb$^{-1}$.

Studies of the strongly interacting EWSBS in $\gamma\gamma$
collision have been considered previously in Refs.~\cite{previous}.
They all concentrate on $\gamma\gamma \rightarrow W_L W_L$ or $Z_L Z_L$.
Unfortunately, the $\gamma\gamma\to W_T W_T$ is almost three orders of
magnitude larger than the $W_LW_L$ signal.
Although we can improve the signal-to-background ratio
by requiring the final state $W$'s away from the beam,
it hardly reduces the $W_TW_T$ background to the level of the $W_LW_L$ signal.
On the other hand, both the $\gamma\gamma\to Z Z$ signal and background
are  absent on tree level.
 But the box diagram contribution to $Z_TZ_T$ has been shown to be very large
at high $m(ZZ)$ region, and so the $Z_T Z_T$ background is
dominant over the $Z_L Z_L$ signal
in the search of the SM Higgs with $m_H\agt 300$ GeV and in probing  the other
strong EWSB signals \cite{ZTZT}.
As illustrated in Refs.~\cite{previous}, the central part of interest is
the $w_L w_L \to w_L w_L$ or $z_Lz_L$, but the effects of the strong EWSBS
only come in on loop level in these processes so that the
effects might  not be so significant.

In the following  we present a new type of processes involving $W_L W_L\to W_L
W_L,\,Z_LZ_L$ at $\gamma\gamma$ colliders, schematically shown in
Fig.~\ref{one} \cite{brodsky}.
These $W_LW_L$ scattering  processes will be in analogy to the
$W_L W_L$ scattering considered at the hadronic supercolliders and
$e^+e^-$ colliders.  The advantages of the processes in Fig.~\ref{one} are that
the $W_LW_L$ scattering is no longer on loop level, and
additional vector bosons  in the final state can be tagged on to eliminate the
large $W_T W_T$ and $Z_T Z_T$ backgrounds.   In addition, both the $W_L^+
W_L^-$ and $W_L^\pm W_L^\pm$ scattering can be studied in $\gamma\gamma$
collision but only one of them can be studied in the $e^+e^-$ or $e^-e^-$
collisions.  Also any $Z_LZ_L$ pair in the final state must come from the
$W_LW_L$ fusion because photon will not couple to $Z$ on tree level.
Totally, we can study four scattering processes,
$W_L^\pm W_L^\pm \to W^\pm_L W^\pm_L$,
$W^+_L W^-_L \to W^+_L W^-_L,\, Z_L Z_L$.

For simplification we will use the effective $W_L$ luminosity inside a photon
in analogy to the effective $W$ approximation.
This approximation will suffice for the purpose here for we will consider the
kinematic region where the EWSBS will interact strongly, or in another words,
in the large invariant mass region of the vector boson pair.
The luminosity function of a
$W_L$ inside a photon in the asymtotic energy limit is given by \cite{zerwas}
\begin{equation}
\label{lum}
f_{W_L/\gamma}(x) = \frac{\alpha}{\pi} \left [ \frac{1-x}{x} +
\frac{x(1-x)}{2}\; \left ( \log \frac{s(1-x)^2}{m_W^2} - 2  \right ) \right
]\,,
\end{equation}
which is in analogy to the luminosity function
$f_{W_L/e}(x) = \frac{\alpha}{4\pi x_{\rm w}} \frac{1-x}{x}$ of $W_L$
inside an electron.
The first term in Eq.~(\ref{lum}) is approximately equal to the luminosity of
$W_L$ inside an electron, and the logarithm factor will enhance the luminosity
at high energy.
This is the reason why the signal rates can be achieved higher than those in
the $e^+e^-$ colliders at the same energy.

\section{Models \& Predictions}

In this section, we will calculate the number of signal events
 predicted by some of the models that have been proposed for the EWSBS. In
$\gamma\gamma$ collision we can study the following subprocesses
\begin{eqnarray}
W_L^+ W_L^- &\rightarrow &  W_L^+ W_L^- \,, Z_L Z_L \;, \\
W_L^\pm W_L^\pm & \rightarrow &  W_L^\pm W_L^\pm \,.
\end{eqnarray}
In analogy to the pion scattering in QCD, the scattering amplitudes of these
processes can be expressed in terms of an amplitude function $A(s,t,u)$.
Their scattering amplitudes are then expressed as
\begin{eqnarray}
{\cal M}(W_L^\pm W_L^\pm \rightarrow W_L^\pm W_L^\pm)&=& A(t,s,u)+A(u,t,s)
\,,\\
{\cal M}( W_L^+ W_L^-  \rightarrow  W_L^+ W_L^- ) &  = & A(s,t,u) + A(t,s,u)
\,,  \\
{\cal M}( W_L^+ W_L^-  \rightarrow  Z_L Z_L ) & = & A(s,t,u) \,,
\end{eqnarray}
up to the symmetry factor of identical particles in the final state.  The
details of each model and the invariant amplitudes predicted by each model
are summarized in Ref.~\cite{bagger}.  Here we only give a brief account of
these models.  The models can be classified
according to the
spin and isospin of the resonance fields, and there  are scalar-like,
vector-like, and nonresonant models.  For scalar-like models we will employ
the standard model with a 1 TeV Higgs, $O(2N)$ model with the cutoff
$\Lambda=2$~TeV, and the model with a chirally-coupled scalar of mass
$m_S=1$~TeV and width $\Gamma_S=350$~GeV.
 For the vector-like models we choose the
chirally-coupled vector field (technirho) of
masses $m_\rho=1$, 1.2, and 1.5~TeV, and $\Gamma_\rho=0.4$, 0.5,
 and 0.6~TeV respectively.
In the case of no light resonances we use the amplitudes predicted by the Low
Energy Theorem (LET) and extrapolate them to high energies.

Each of the $W_L W_L$ scattering amplitudes grows with energy until reaching
the resonances, {\it e.g.} SM Higgs boson of the minimal SM.  The presence of
the resonances (scalar or vector) is the natural unitarization to the
scattering amplitudes, except there might be slight violation of unitarity
around the resonance peak.  After the resonance, the scattering amplitudes
will stay below the unitarity limit.
But for the nonresonant models the unitarity
is likely to be saturated before reaching the lightest resonance.  Here we
employ the LET amplitude function, $A(s,t,u)=s/v^2$, for the nonresonant
models.  From the partial
wave analysis, the only nonzero partial wave coefficients
$a^I_J$ are $a^0_0$, $a^1_1$, and $a^2_0$.  Among the nonzero $a^I_J$'s,
$a^0_0$ saturates the unitarity ($|a^I_J|<1$) at the lowest energy
$4\sqrt{\pi}v\approx 1.7$~TeV.  So for the $\gamma\gamma$ colliders
of 1.5 TeV, unitarity violation should not be a problem, therefore, we simply
extrapolate the LET amplitudes without any unitarization.

We show the number of signal events predicted by these models for each
scattering channel  in
Table~\ref{table1}, with $\sqrt{s_{\gamma\gamma}}=1.5$~TeV
 and integrated luminosity of 100~fb$^{-1}$, and under the
acceptance cuts of
\begin{equation}
M_{WW}\;{\rm or}\; M_{ZZ} > 500\;{\rm GeV}\quad {\rm and}\quad |y(W,Z)|<1.5\,.
\end{equation}
One interesting thing to note here is that different channel is sensitive to
different new physics.  If the underlying dynamics of the EWSBS is scalar-like
the signal is more likely to be found in the $W_L^+ W_L^-$ channel, and next
at the $Z_LZ_L$ channel, due to the presence of
$I=0,J=0$ scalars.  But if the underlying
dynamics is vector-like the signal in the $W_L^+W_L^-$ channel will be far
more important that the $Z_L Z_L$ channel.  On the other hand, if no light
resonances are within reach the $Z_LZ_L$ channel has the largest signal rate,
and next
is the $W^\pm_L W^\pm_L$ channels.  So by counting the number of $W^\pm_L
W^\pm_L$, $W^+_L W^-_L$, and $Z_L Z_L$ pairs
in the final state one can tell the
different structure of the EWSBS \cite{han}.  But to distinguish a $W$ from a
$Z$ by the dijet mass measurement is not a trivial issue, though we can use
the $B$-tagging to distinguish a $W$ from a $Z$ somehow.  For a discussion on
this subject please see, {\it e.g.}, Ref.~\cite{han}.

The number of signal events in Table~\ref{table1} does not include any
detection efficiencies of the $W_L$'s coming out from the
strong scattering region, nor the tagging efficiencies for the spectator $W$'s.
The tagging efficiencies for the spectator $W$'s will be dealt with in the next
section.  The detection efficiencies of the $W_L$'s consist of the branching
ratios of the $W_L$ into jets or leptons, and the tagging efficiencies of
these decay products.  The branching ratio BR($W\to jj)\approx$BR$(Z\to jj)
\approx 0.7$.
Assuming a 30\% (reasonable to pessimistic)
tagging efficiencies for the decay products, we have about
15\% overall detection efficiencies for the $W_LW_L$ coming out from the
strong scattering region.

\section{Tagging the Spectator $W$'s}

So far we have not considered any backgrounds nor background suppression
techniques. In our calculation, we use the effective $W_L$ luminosity which
does not predict the correct kinematics for the spectator $W$'s, and therefore
any acceptance cuts on the spectator $W$'s will be unrealistic.  However, we
need to tag at least one  or both of these spectator $W$'s in order to
eliminate the
enormous $\gamma\gamma\to W_T W_T,\,Z_T Z_T$ backgrounds.  One way to remedy
is to carry out an exact SM calculation of $\gamma\gamma\to WWWW$ or $WWZZ$
with a heavy Higgs boson, and estimate the acceptance efficiencies on tagging
the spectator $W$'s, and then apply these efficiencies to the other models
which  can only be calculated using the effective $W_L$ luminosity.

However, the calculations of the processes
$\gamma\gamma\to WWWW$ or $WWZZ$ are
non-trivial.  Instead, we can obtain the tagging efficiencies by calculating  a
simpler process $\gamma\gamma \to WWH$ for $m_H\approx $1 TeV, with and
without imposing acceptance cuts on the final state $W$'s.  We will calculate
the total cross section for $\gamma\gamma\to WWH$ without any cuts, and also
the cross section with the acceptance cuts
\begin{equation}
p_T(W) > 25\;{\rm GeV},\qquad |y(W)| < 1.5\;{\rm or}\;2
\end{equation}
on either one or both of the $W$'s.  The cross sections are presented in
Table~\ref{table2} for $m_H=1$~TeV.  There are two tagging efficiencies
corresponding to tagging at least  one or both of the spectator $W$'s.
{}From Table~\ref{table2}, if we require the spectator $W$'s within a rapidity
of $|y(W)|<1.5$ the tagging efficiencies are 91\% and 42\% for tagging at
least one or both the $W$'s respectively.  To eliminate the $W_T W_T$ or $Z_T
Z_T$ backgrounds we need only tag one of the spectator $W$'s  plus the
$W_L$'s from the strong scattering.  A further confirmation by tagging
two spectator $W$'s
will result in an efficiency of only 42\%.  But if we tag both spectator $W$'s
within the rapidity $|y(W)|<2$ the double-tag efficiency increases to 82\%.
This drastic difference of the double-tag efficiencies between rapidity cut of
1.5 and 2 demonstrates that it is likely (40\% chance) to have at least one
spectator $W$ in the forward rapidity region $1.5<|y(W)|<2$.
Next we can multiply these efficiencies to the numbers in
Table~\ref{table1} to get a more
reliable number of signal events when the spectator $W$'s are tagged.
Taking into account of the 15\% (from the last section) detection efficiency
for the $W_L W_L$ plus the tagging efficiency of at least one or both of the
spectator $W$'s, we still have at least 10\% overall efficiency.  With 10\%
efficiency we still have a sizeable number of signal events.
Scalar-type  models will be shown up in the $W^+W^-\to W^+W^-$ channel with
at least 47 events.   The vector-like models will also be shown up in the
$W^+W^-\to W^+W^-$ channel if the vector resonance is within reach of the
energy of the $\gamma\gamma$ collider.   For nonresonant models we  have
about 15 events for the $W^\pm W^\pm \to W^\pm W^\pm$ channels
and 17 events for $W^+W^-\to ZZ$ channel.

\section{Background Discussions}

The continuum productions of $\gamma\gamma\to WWWW$ and $WWZZ$, together with
the heavy quark production of
$\gamma\gamma\to t\bar t t\bar t$ followed by  the top decays into
$W$'s, form the irreducible set of backgrounds.  They are the SM
predictions that any significant excess of $W_L W_L$ or $Z_L Z_L$
events will indicate some
kinds of new physics for the EWSBS.  The other reducible backgrounds include
the productions of $W$'s with jets, $Z$'s with jets, and multi-jet.

The $WWWW$ and $WWZZ$ productions are of order $\alpha_w^4$, and so should be
at most the same level as our strong $W_L W_L$ signal.  Although
the $t\bar t t\bar t$  background is ${\cal O}(\alpha_s^2/\alpha_w^2)$ larger
 than the $WWWW$  background, we can to certain extent reduce it by
reconstructing the top and by imposing the top-mass constraints.  The other
QCD backgrounds of $W$'s or $Z$'s with jets are reducible by the $W$ or $Z$
mass constraints.

In addition, we can make use of the kinematics of the spectator $W$'s
and the strongly scattered $W_L$'s \cite{kingman}.  The $p_T$ of the spectator
$W$'s should be of order $m_W/2$ after the photon emits an almost on-shell
$W_L$, which then participates in the strong scattering.
Also, as mentioned in the last section,
at least one of the spectator $W$'s tend to go forward in the rapidity
region  $|y(W)|>1.5$.
On the other hand, the $W_L W_L$ after
the strong scattering come out in the  central rapidity region
 with large  $p_T$ and  large invariant mass, and  back-to-back in the
transverse plane, which are all due to the strong interaction of the EWSBS.
But it is hardly true for the backgrounds.  Acceptance cuts can be formulated
based on the above arguments to substantially reduce the backgrounds
\cite{future}.

In conclusions, we have demonstrated another type of
 processes in $\gamma\gamma$
collision that can probe the strongly interacting EWSBS scenario.  The
processes do not involve the indirect loop effects, and also are safe from the
huge $W_T W_T$ or $Z_T Z_T$ backgrounds due to the presence of the
spectator $W$'s.  Even with
only 10\% overall efficiency we still have enough signal events with
100~fb$^{-1}$ luminosities.  Irreducible backgrounds from $WWWW$, $WWZZ$, and
$t\bar t t\bar t$ can be reduced by considering the special kinematics of the
strongly scattered $W_L$'s and the spectator $W$'s.  Other reducible
backgrounds are  reduced by the mass constraints.

\acknowledgements

We are grateful to V.~Barger, D.~Bowser-Chao, and T.~Han for useful
discussions. This work was supported by the U.~S. Department of Energy,
Division of High Energy Physics, under Grant DE-FG02-91-ER40684.


\begin{table}
\caption{\label{table1}
The number of the signal events for the strong $W_L W_L$ scattering
predicted by various models at $\gamma\gamma$ collider of
$\sqrt{s}=1.5$~TeV. The acceptance cuts on the final
$W_LW_L$ or $Z_LZ_L$ are: $m(WW,ZZ)>500$~GeV and $|y(W,Z)|<1.5$.  The
luminosity is assumed 100~fb$^{-1}$. No efficiencies are included here.
}
\begin{tabular}{lccc}
 & $W^+_L W^+_L \to W^+_L W^+_L$ &  $W^+_L W^-_L \to W^+_L W^-_L$ &
  $W^+_L W^-_L \to Z_L Z_L$ \\
\hline
(1) SM Higgs  &  & & \\
\mbox{} $m_H=1$ TeV & 88 & 1600  & 760\\
(2) chirally-coupled scalar &&& \\
\mbox{} $m_S=1$ TeV, $\Gamma_S=350$ GeV & 100 & 570 &  430\\
(3) O(2N) & 90 & 470 & 350\\
\hline
(4) chirally-coupled vector &&& \\
 a. $m_V=1$ TeV, $\Gamma_V=0.4$ TeV  & 180 & 2400 & 280\\
 b. $m_V=1.2$ TeV, $\Gamma_V=0.5$ TeV  & 52  & 590  & 29 \\
 c. $m_V=1.5$ TeV, $\Gamma_V=0.6$ TeV  & 88 & 120 & 40 \\
\hline
LET & 150 & 110 & 170 \\
\end{tabular}
\end{table}

\begin{table}
\caption{\label{table2}
Table showing the cross sections (fb) for the process
$\gamma\gamma\to WWH$ with a SM Higgs boson of mass $m_H=1$~TeV at
$\sqrt{s_{\gamma\gamma}}=1.5$~TeV,
with and without imposing acceptance cuts on the
final state $W$'s.  The acceptance cuts are $p_T(W)>25$~GeV and $|y(W)|<1.5$ or
2.  The second column shows the total cross section without cuts.
The third column corresponds to tagging at least one of the $W$'s, and the
last column corresponds to tagging both.  The percentages in the parentheses
are the efficiencies.
}
\begin{tabular}{lccc}
$|y(W)|<$    &  No cuts   &  Tagging at least one $W$   & Tagging both $W$'s \\
\hline
 -           &    14.7      &       -             & - \\
1.5          &     -        &      13.4 (91\%)    & 6.16 (42\%) \\
2.0          &     -        &      14.5 (98.5\%)  & 12.1 (82\%) \\
\end{tabular}
\end{table}
\figure{\label{one}
Schematic diagram for the $W_L W_L$ scattering in $\gamma\gamma$ collision.
}

\end{document}